\documentclass[preprint,showkeys,secnumarabic,amsmath,amssymb,nofootinbib]{revtex4}
\usepackage{graphicx}

\begin{document}

\title{Stability Analysis of DGP Brane-world Model with Agegraphic Dark Energy}

\author{A. Ravanpak}
\email{a.ravanpak@vru.ac.ir} \affiliation{Department of Physics, Vali-e-Asr University of Rafsanjan, Rafsanjan, Iran}
\author{G. F. Fadakar}
\email{g.farpour@vru.ac.ir} \affiliation{Department of Physics, Vali-e-Asr University of Rafsanjan, Rafsanjan, Iran}

\date{\small {\today}}

\begin{abstract}
The aim of this work is to apply the dynamical system approach to study the linear dynamics of the normal DGP brane-world model with agegraphic dark energy. The stability analysis of the model will be investigated and the phase plane portrait will be illustrated. The nature of critical points will be analyzed by evaluating the eigenvalues of a linearized Jacobi matrix. Also, the statefinder diagnostic procedure will be applied to show the slight deviation of our model from the $\Lambda$CDM model. One of the most interesting results of this work is the great alleviation of the coincidence problem.
\end{abstract}

\keywords{dynamical system, stability, DGP, agegraphic dark energy, coincidence problem}

\maketitle

\section{Introduction}\label{s:1}

We know from observations that our Universe is experiencing an accelerated expansion phase \cite{Riess,Perlmutter}. This acceleration can be described in two distinct scenarios. In one of them, one can modify the left-hand side of Einstein's equation, which is called \textit{modified gravity} \cite{Nojiri,Capozziello}, and in the other, one can add a component with negative pressure to the right-hand side of Einstein's equation, which is dubbed \textit{dark energy} (DE). A great variety of DE models have been proposed in the literature \cite{Sahni,Caldwell,Mukhanov,Padmanabhan,Elizadle,Kamenshchik,Bento}. Among them holographic dark energy (HDE) \cite{Li}, and agegraphic dark energy (ADE) \cite{Cai} which are based on the holographic principle and the quantum fluctuations of space-time, respectively, are of particular interest because they contain some important features of a quantum gravity theory, although, a complete and comprehensive formulation of this theory has not yet been established.

In the ADE approach, K\'{a}rolyh\'{a}zy and his collaborators showed that in Minkowaskian space-time the distance $t$, cannot be known to a better accuracy than \cite{Karolyhazy}
\begin{equation}
   \delta t=\gamma t_p^{2/3}t^{1/3}\,,
\end{equation}
where $\gamma$, is a dimensionless constant of order unity and $t_p$, denotes the reduced Planck time. Then, the authors in \cite{Maziashvili}, using the time-energy uncertainty relation estimated the quantum energy density of the metric fluctuations of Minkowaskian space-time where it can be considered that the energy density of ADE as
\begin{equation}\label{rho}
\rho_{DE}\sim\frac{1}{t_p^2t^2}\sim\frac{M_p^2}{t^2}\,,
\end{equation}
where $M_p$, represents the reduced Planck mass. Replacing the proper time scale $t$, in the above equation with the age of the Universe, $T=\int_0^a\frac{da}{Ha}$, where $a$ and $H$, are respectively the scale factor and the Hubble parameter, Cai obtained the energy density of ADE as \cite{Cai}
\begin{equation}\label{rho2}
\rho_{DE}=\frac{3n^2M_p^2}{T^2}\,.
\end{equation}
Here, $3n^2$, is a numerical factor that parameterizes some uncertainties such as the effects of curved
space-time\footnote{Because the energy density is derived for Minkowskian space-time}, and the species of quantum fields in the Universe.

On the other hand, extra dimensional theories have attracted a considerable amount of attention in the past two decades \cite{Randall,Bouhmadi,Bouhmadi2,Lee,Shtanov,Zhang}. In these models, our four-dimensional (4D) Universe is a  brane embedded in a higher dimensional space-time called bulk, where the standard model of particle physics is confined to the brane and only the graviton can propagate into the bulk. The extra dimensions affect the Friedmann equations on the brane by inducing a few additional terms \cite{Binetruy,Binetruy2,Shiromizu}. In particular, the brane-world model proposed by Dvali, Gabadadze and Porrati (DGP), in which the bulk is an infinite five-dimensional (5D) Minkowski space-time, has been studied, recently \cite{Dvali}. This model has two branches (solutions), depending on how the brane can be embedded in the bulk: the self-accelerating branch that explains the late time acceleration of the Universe without any DE ingredient but suffers from ghost instability, and the normal branch which needs a DE component to produce an accelerated expansion phase (but does not have the ghost problem).

Apart from the above subjects, mathematicians use stability theory to address the stability of solutions of differential equations under small perturbations. Dynamical system techniques have greatly been used in studying cosmological models \cite{Wainwright,Coley,Zonunmawia,Biswas,Zhang2,Quiros,Ravanpak2,Zhang3,Nozari,Arvin,Chimento}. Their main advantage is the possibility of studying all solutions with admissible initial conditions. As there are always some uncertainties in the initial conditions of a model, a physically meaningful mathematical model that presents detailed information on the possible deviations of trajectories of the dynamical system from a specified reference trajectory is very useful. For instance, in \cite{Zonunmawia}, the authors have investigated a global dynamical system perspective of the cosmological dynamics of brane gravity and found that important cosmological behaviors are consistent with brane gravity.

The incorporation of the DGP brane-world model and different kinds of DE components such as the cosmological constant, a scalar field, Chaplygin gas, HDE and ADE, has been discussed in the literature \cite{Ravanpak,Chimento,Bouhmadi3,Wu,Farajollahi}. In most cases, the dynamical system approach has been studied in detail. In \cite{Zhang2}, the stability of the Einstein static universe in the context of DGP brane-world gravity has been investigated. Also, in \cite{Quiros} and \cite{Ravanpak2}, the stability analysis of a DGP model has been studied in the presence of a quintessence scalar field and a tachyon scalar field, respectively, and it has been shown that the dynamics of the DGP model can be very rich and complex. In \cite{Quiros}, the authors have indicated that depending on the type of quintessence scalar field potential, the dynamical screening of the scalar field energy density that is a phenomenon of 5D nature could be a generic solution. Also, they have claimed that matter-scaling solutions could exist, and even could be an attractor. On the other hand, in \cite{Ravanpak2}, we have found critical submanifolds that indicate the effect of an extra dimension, as well as an interesting late time transition of the universe from accelerated expansion to decelerated expansion. In \cite{Zhang3}, using the dynamical system procedure the authors have indicated the phantom divide crossing in the DGP model not in the presence of a phantom scalar field but considering just an ordinary quintessence scalar field. Also, they have claimed that this model does not have a matter-scaling solution. In \cite{Nozari}, the authors have extended the procedure in \cite{Zhang3}, to the non-minimal coupling situation, and in the presence of quintessence and phantom scalar fields, separately. They have shown the late time cosmic acceleration and also phantom divide crossing in some cases. In \cite{Arvin,Chimento}, the authors have indicated the avoidance of big-rip singularity in an agegraphic DGP model and a quintessence DGP model, respectively, although, in \cite{Chimento}, phantom crossing does occur.

In this manuscript we consider the normal branch of the DGP model in the presence of ADE and investigate the stability analysis of the model. Our main motivation in this mixed model is the quantum nature of its components, because ADE comes from quantum gravity theory, and the extra dimensional gravity is a result of string theory. In \cite{Arvin}, we have studied the same model but in the presence of interaction between the dark sectors of the Universe. However, as the interaction term is estimated to have a very small value, here, we address the consequences of ignoring that interaction. We will find that the results are different from the ones in \cite{Arvin}. Also, we use the statefinder diagnostic to compare our model with the standard $\Lambda$CDM model. We will show numerically that the trajectory of the Universe in the respective phase plane passes through the critical points of the model and the point related to the $\Lambda$CDM model, as well. In addition, we will investigate the capability of the model to solve the coincidence problem.

The paper is organized as follows: In Sec.\ref{s:2}, we construct the model, introduce our new variables and write respective ordinary differential equations. The critical points and related eigenvalues will be discussed in this section, as well. Sec.\ref{s:3}, deals with the statefinder diagnostic approach. The coincidence problem is also studied in this section. Sec.\ref{s:4}, includes a summary and the conclusion of the article.

\section{The Model and Stability Analysis}\label{s:2}

In this section, firstly, we introduce a normal DGP brane-world model in the presence of ADE and then utilize the dynamical system approach to investigate the model, carefully. We start with the cosmological equations of the model. Assuming the brane is spatially flat, homogeneous and isotropic, the Friedmann equation of the brane can be written as \cite{Deffayet}
\begin{equation}\label{fried}
H^2+\frac{H}{r_c}=\frac{1}{3M_p^2}(\rho_m+\rho_{DE})\,,
\end{equation}
where $\rho_{m}$, is the energy density of the matter content of the Universe that is dominant by dark matter (DM), and $r_c$, is called the crossover distance which determines the transition from the 4D to 5D regime. From the conservation equations of DM and DE on the brane we have
\begin{equation}\label{conservationDM}
    \dot\rho_m +3H \rho_m = 0\,,
\end{equation}
\begin{equation}\label{conservationDE}
    \dot\rho_{DE} +3H \rho_{DE}(1+\omega_{DE}) = 0\,.
\end{equation}
Differentiating Eq.\eqref{rho2}, we obtain
\begin{equation}\label{rhodotDE}
    \dot\rho_{DE}=-2H\rho_{DE}\frac{\sqrt{\Omega_{DE}}}{n}\,,
\end{equation}
in which
\begin{equation}\label{ODE}
    \Omega_{DE}=\frac{\rho_{DE}}{3M_p^2H^2}=\frac{n^2}{H^2T^2}\,.
\end{equation}
Inserting Eq.(\ref{rhodotDE}), into Eq.(\ref{conservationDE}), we find the equation of state (EoS) parameter of ADE, as
\begin{equation}\label{eos}
\omega_{DE}=-1+\frac{2}{3n}\sqrt{\Omega_{DE}}\,,
\end{equation}
which indicates that the EoS parameter of ADE, never crosses the phantom divide. Also, the Raychaudhuri equation of our model can be obtained using Eqs.(\ref{fried}), (\ref{conservationDM}) and (\ref{rhodotDE}), as
\begin{equation}\label{Raychaudhuri}
    \dot H=\frac{-\rho_m-\frac{2}{3n}\rho_{DE}\sqrt{\Omega_{DE}}}{M_p^2(2+\frac{1}{Hr_c})}\,.
\end{equation}

To apply dynamical system analysis to the model, one has to introduce some auxiliary variables to transform the cosmological equations of motion into a self-autonomous dynamical system. Here, we introduce the following dimensionless phase variables:
\begin{eqnarray}\label{nv}
 x&=&\sqrt{\frac{\rho_m}{3M_p^2(H^2+\frac{H}{r_c})}},\nonumber\\ y&=&\sqrt{\frac{\rho_{DE}}{3M_p^2(H^2+\frac{H}{r_c})}},\nonumber\\ z&=&\sqrt{1+\frac{1}{Hr_c}}
\end{eqnarray}
which show standard 4D behavior for $r_c\rightarrow \infty$. The new variable $z$, satisfies $z\geq1$, as $H$ and $r_c$, have positive values. Also, the Friedmann equation, Eq.(\ref{fried}), yields the constraint
\begin{equation}\label{constraint}
x^2+y^2=1,
\end{equation}
which means that our system has only two degrees of freedom. With attention to this constraint and because our phase variables cannot be negative, they have to satisfy the constraints $0 \leq x\leq1$, and $0 \leq y\leq1$.

On the other hand, we can rewrite the Raychaudhury equation in terms of the new variables as
\begin{equation}\label{ray}
\frac{\dot H}{H^2}=\frac{-3z^2(x^2+\frac{2}{3n}y^3z)}{z^2+1}\,.
\end{equation}
In addition, the EoS parameter of ADE, Eq.(\ref{eos}) and the total EoS parameter of the Universe can be written as
\begin{equation}\label{eos2}
    \omega_{DE}=-1+\frac{2}{3n}yz\,,
\end{equation}
and
\begin{equation}\label{eostot}
    \omega_{tot}=\frac{\omega_{DE}\rho_{DE}}{\rho_m+\rho_{DE}}=-1+x^2+\frac{2}{3n}y^3z\,,
\end{equation}
respectively. With attention to the phase-space variables, Eq.(\ref{nv}), and the Friedmann constraint, Eq.(\ref{constraint}), and also Eq.(\ref{ray}), we reach to the following autonomous system of ordinary differential equations in $(y,z)$ space
\begin{eqnarray}\label{ds}
  y' &=& -\frac{y^2z}{n}+\frac y2\left(3(1-y^2)+\frac{2}{n}y^3z\right)\,, \\
   z' &=& \frac{z(z^2-1)}{2(z^2+1)}\left(3(1-y^2)+\frac{2}{n}y^3z\right)\,,
\end{eqnarray}
in which the prime means derivative with respect to $\ln a$. These equations interpret the evolution of the DGP brane-world model with non-interacting DM and ADE.

Stability analysis studies the behavior of a system in the vicinity of its critical points. To determine the critical points of the dynamical system above we must impose the conditions $y'=0$, and $z'=0$, simultaneously. The admissible results that satisfy the limitations on $y$, and $z$, and also the Friedmann constraint, Eq.(\ref{constraint}), are shown in Table \ref{tbl:1}.

\begin{table}[h]
 \caption{The fixed points of the normal DGP model with ADE.} 
  \label{tbl:1}
 \begin{tabular}{|c|c|c|c|c|c|} \hline
  & $(z,y)$& eigenvalues&$\omega_{tot}$& description&stability \\ \hline\hline
 $A$ & (1 , 0)&(3/2 , 3/2)& 0 & DM domination & unstable \\ \hline
 $B$ & (1 , 1)&$(1/n ,(2-3n)/n)$& $-1+2/(3n)$& DE domination & saddle\\
 \hline
 \end{tabular}
\end{table}

The point $A$, relates to a matter dominated era because with attention to the Friedmann constraint, $y=0$, is equivalent to $x=1$. On the other hand, using Eq.(\ref{nv}), we have $x=\sqrt{\Omega_m}/z$, where $\Omega_m=\rho_m/3M_p^2H^2$. Because at point $A$, we have $z=1$, so this point shows a matter dominated regime. In the same way, point $B$, is related to a DE dominated era. Thus, there does not exist a matter-scaling solution (in which the energy density of DM and ADE are proportional) in our model. This result is similar to the case in \cite{Quiros}, for a self-interacting scalar field with a constant potential in DGP cosmology.

\begin{figure}[h]
    \centering
    \includegraphics[width=0.45\textwidth]{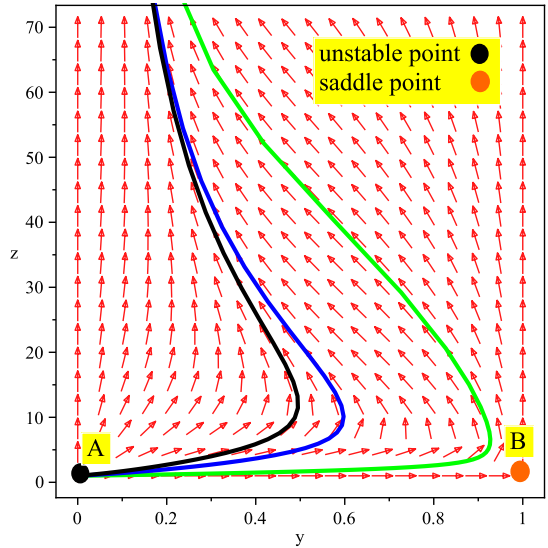}
    \caption{The 2D phase plane corresponding to the critical points.}\label{fig:field}
\end{figure}

We should note here that if we evaluate Eq.(\ref{eostot}) at point $B$, we find a necessary condition for the parameter $n$, as $n>2$, to guarantee the late time acceleration in the normal branch of DGP model, $w_{tot}<-2/3$ \cite{Gumjudpai}. Inserting the constraint $n>2$, into the eigenvalues of point $B$, we find that point $B$, is always a saddle point. In \cite{Arvin}, we have shown that in the same model but in the presence of interaction, the critical point related to a DE dominated era will be a stable critical point. Fig.\ref{fig:field}, demonstrates the phase portrait of our dynamical system in the $yz$-plane for $n=4$.

We should note some important results here. One can rewrite the Friedmann equation, Eq.(\ref{fried}) as
 \begin{equation}\label{refried}
    \Omega_m+\Omega_{DE}=z^2=1+\frac{1}{Hr_c}.
\end{equation}
On the other hand from Table \ref{tbl:1}, one can see that at both critical points we have $z=1$, which is related to the formal limit $r_c\rightarrow\infty$, which indicates that these points correspond to standard 4D behavior where $\Omega_m+\Omega_{DE}=1$. Therefore the line $z=1$ in Fig.\ref{fig:field}, is the trajectory of a pure 4D cosmology that arises from any initial condition as $(y,z=1)$. It shows that the 4D Universe starts in a matter dominated epoch and approaches a DE dominated era, as we expect in standard 4D cosmology. In this case we deal with a phase line where critical points $A$ and $B$, are its end points. So, the matter dominated solution, point $A$, and the DE dominated solution, point $B$, will necessarily be the repeller and attractor, respectively. All other phase trajectories in Fig.\ref{fig:field}, which leave the phase line $z=1$ and probe the phase plane $(y,z)$, arise from any other initial condition with $z\neq1$ and show the effects of an extra dimension in our model. It is in this case that point $B$, is a saddle point. In all these cases starting from a matter dominated era the Universe no longer experiences a 4D DE dominated period because as from Eq.(\ref{refried}) another term will be dominated which we can call it $\Omega_{DGP}=\Omega_{DE}-1/Hr_c$. Although one can consider it as an effective DE term, it clearly differs from the one related to the saddle point $B$. It can be seen in Fig.\ref{fig:field} that all these trajectories are repelled from the line $y=1$. Here, as in the case in the presence of interaction \cite{Arvin}, we do not find a critical submanifold. But with attention to Fig.\ref{fig:field}, as the case for a quintessence scalar field with a constant potential in DGP cosmology \cite{Quiros}, gravitational screening is a solution in our model but apparently only at infinity.

\begin{figure}[h]
    \centering
    \includegraphics[width=0.45\textwidth]{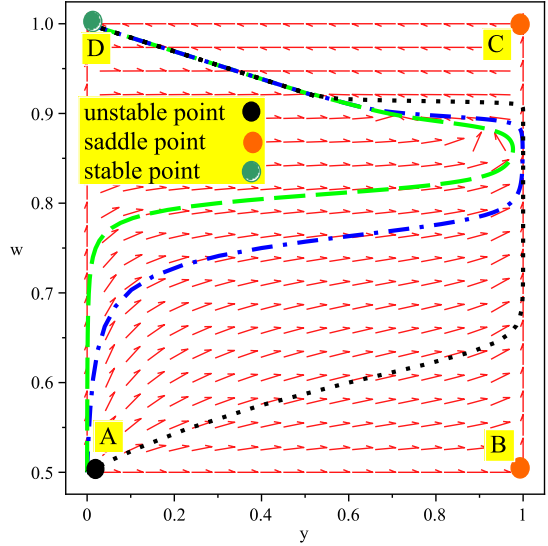}
    \caption{The 2D phase plane corresponding to the new critical points.}\label{fig:field2}
\end{figure}

For analyzing the behavior of trajectories and finding the critical points at infinity we condense our dynamical system defining a new variable $w$ as
\begin{equation}\label{newvariable}
    w=\frac{z}{1+z}
\end{equation}
such that for $z=1$, we find $w=1/2$, and in the limit $z\rightarrow\infty$ we get $w=1$. So, the new variable is bounded as $1/2\leq w\leq1$. The new autonomous system of ordinary differential equations is obtained as
\begin{eqnarray}\label{ds2}
  \frac{dy}{d\lambda} &=& -\frac{y^2w}{n}+\frac{y}{2}\left(3(1-y^2)(1-w)+\frac{2}{n}y^3w\right)\,, \\
   \frac{dw}{d\lambda} &=& \frac{w(2w-1)(1-w)^3}{2(2w^2-2w+1)^3}\left(3(1-y^2)(1-w)+\frac{2}{n}y^3w\right)\,,
\end{eqnarray}
in which $\frac{d}{d\lambda}\equiv(1-w)\frac{d}{d\ln a}$, is utilized to remove the infinities for $w=1$, in the new system. We find that there are four physical critical points for the system above, i.e., those that satisfy the constraints $0\leq y\leq1$ and $1/2\leq w\leq1$. Two of them are $(y=0,w=1/2)$ and $(y=1,w=1/2)$, with eigenvalues $(\frac{3}{4},\frac{3}{4})$ and $(\frac{2-3n}{2n},\frac{1}{2n})$, respectively. The former is an unstable point and the latter, considering the constraint $n>2$, is a saddle point. One can easily check that the first one is related to the critical point $A$, and the second one to the critical point $B$, in Table \ref{tbl:1}. The third critical point is $C\doteq(y=1,w=1)$, with eigenvalues $(0,\frac{2}{n})$, and the fourth is $D\doteq(y=0,w=1)$ with eigenvalues $(0,0)$. Since in both of the latter cases there is at least one zero eigenvalue, we are required to apply some methods such as the center manifold theory \cite{Perko}, rather than the linear approximation method, to investigate their stability. But, since this analytical study lies beyond the scope of the present work, we turn to the numerical approach. From Fig.\ref{fig:field2}, we conclude that $C$, is a saddle point but $D$, is an attractor solution.

\section{Features of the Model}\label{s:3}

\subsection{Statefinder Diagnostic}\label{ss:1}

The statefinder diagnostic is one of the most practical techniques and is a reliable approach to discriminate between various DE models. It is based on a pair of new geometrical variables that are related to the third derivative of the scale factor with respect to time. In a flat Universe the statefinder parameters are defined as \cite{Sahni3}
\begin{equation}\label{rs}
    r=\frac{\dddot a}{aH^3},\quad s=\frac{r-1}{3(q-\frac{1}{2})}\,,
\end{equation}
Here, $q$, is the deceleration parameter. To classify different DE scenarios one can compare the respective trajectories in the $rs$-plane. Also, the deviation from the $\Lambda$CDM model which is expressed by the point $(r=1, s=0)$, can be studied in this way.

\begin{figure}
    \centering
    \includegraphics[width=0.42\textwidth]{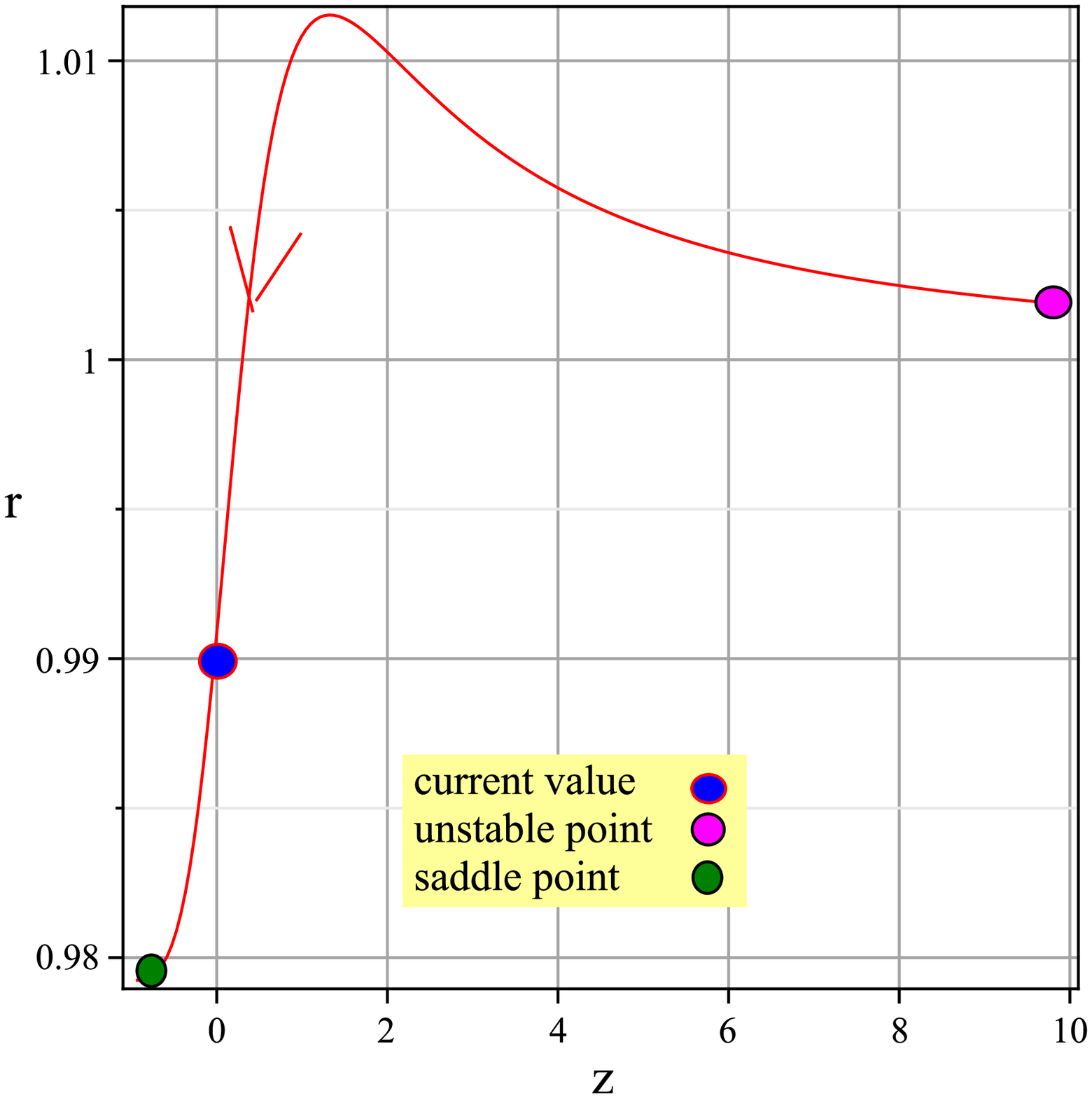}
    \includegraphics[width=0.42\textwidth]{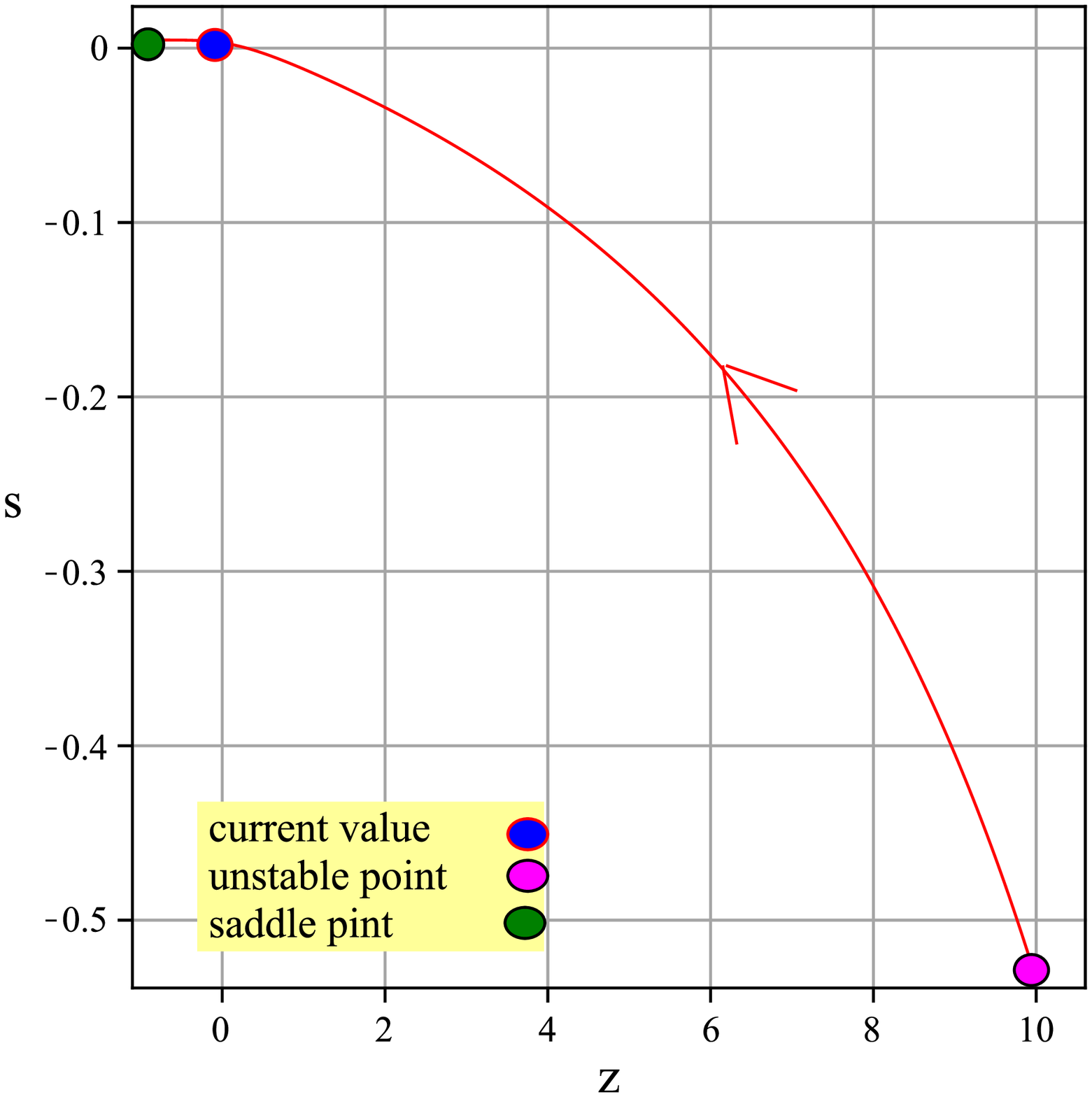}
    \includegraphics[width=0.42\textwidth]{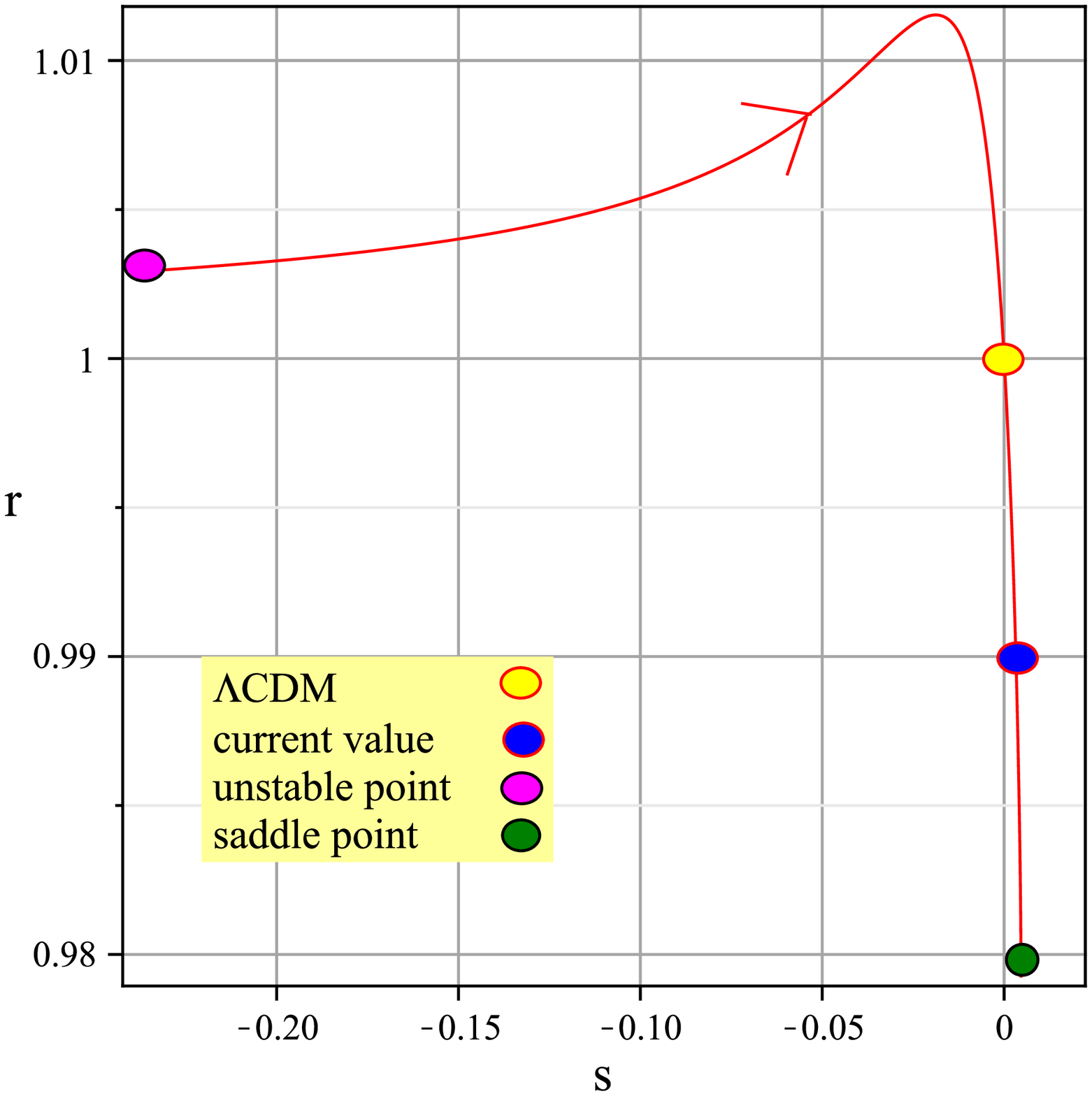}
    \caption{The plot of statefinder parameters $r$ and $s$, versus the redshift parameter $z$, and also the trajectory in the $rs$-phase plane. We have used the initial conditions $H_0=0.67$ and $\Omega_{m0}=0.31$ from Planck Collaboration's results \cite{Ade}, and also $n=2$ and $\Omega_{rc}=1/4r_c^2H_0^2=0.002$.}\label{fig:rs}
\end{figure}

From Fig.\ref{fig:rs}, which shows the phase trajectory in the $rs$-plane, it is clear that the Universe leaves an
unstable state in the past, passes through the point related to the $\Lambda$CDM model and finally approaches the saddle state in the future. Also, the point of the current value is very close to $(r=1, s=0)$. So, from the point of view of the statefinder diagnostic, our model is in good agreement with the $\Lambda$CDM model. On the other hand, in two distinct graphs we have illustrated variation of $r$ and $s$, with respect to the redshift parameter. One can find from Fig.\ref{fig:rs}, that $s$, is increasing in all history, but $r$, has experienced a change from increasing to decreasing in the near past, though it did not vary significantly.

\subsection{Coincidence Problem}\label{ss:2}

The coincidence problem which is a fine-tuning cosmological problem appears because the energy densities of DM and DE are of the same order at the present time. Many authors have used different approaches to solve or at least alleviate this problem \cite{Amendola,Amendola2,Chimento2,Egan,Karwan,Olivares,campo,campo2,Scherrer,Malquarti,Picon,Picon2}. Some of them utilize the attractor solution and try to show that the ratio between the dark sectors, $R\equiv\rho_m/\rho_{DE}$, is independent of the initial conditions \cite{Amendola,Amendola2,Karwan}. Another group demonstrates that $R$, does not vary much during the evolution of the Universe \cite{campo,campo2}, and some others indicate that $R$, approaches a constant value at late times or changes slower than the scale factor today \cite{Amendola2,Chimento2}. Some of them employ non-interacting dynamical DE models \cite{Scherrer,Malquarti,Picon,Picon2}, but others consider the models in which an interaction between the dark sectors of the Universe is taken into account \cite{Amendola,Amendola2,Chimento2,Karwan}.

In an earlier work we tried to show the role of an extra dimension in resolving this problem in a normal DGP scenario in the presence of cosmological constant, $\Lambda$, as the DE component \cite{Ravanpak}. Here, we are going to demonstrate the effect of combining the normal DGP model with ADE, in addressing the coincidence problem. Comparing Eq.(\ref{fried}), with the standard Friedmann equation, we can introduce an effective DE component in our model with the energy density $\rho_{eff}$, as
\begin{equation}\label{omega_eff}
\rho_{eff}=\rho_{DE}-\frac{3M_p^2H}{r_c}\,.
\end{equation}
Then, we can represent the ratio between the dark sectors in our model as $R\equiv\rho_m/\rho_{eff}$.

\begin{figure}
\centering
\includegraphics[width=0.45\textwidth]{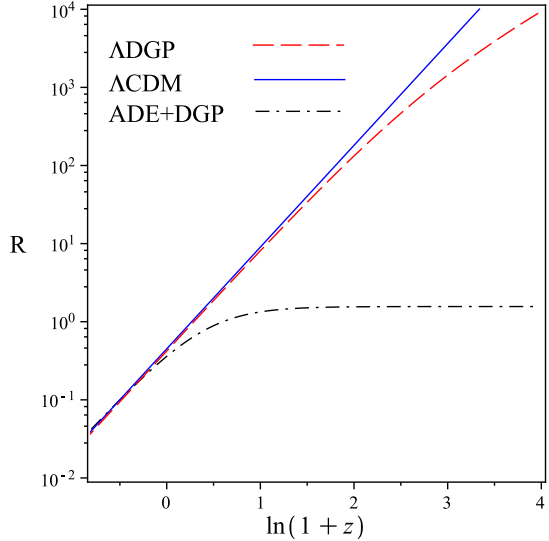}
\caption{The coincidence problem is significantly alleviated in the ADE+DGP model. We have used the initial conditions $H_0=0.67$ and $\Omega_{m0}=0.31$ from the Planck Collaboration's results \cite{Ade}, and also $n=2$ and $\Omega_{rc}=1/4r_c^2H_0^2=0.002$. The parameter $z$ in $\ln(1+z)$, is the redshift.}\label{fig:coin}
\end{figure}

Fig.\ref{fig:coin}, illustrates the behavior of $R$, in our model and compares it with the one of the standard $\Lambda$CDM model and the one in the $\Lambda$DGP model. One of the initial values we have used to plot Fig.\ref{fig:coin}, is $\Omega_{rc}=1/4r_c^2H_0^2=0.002$. Obviously, with this choice, $r_c$ has a finite value and so $z\neq1$. Therefore the effect of the extra dimension has surely been imported in this figure. Although Fig.\ref{fig:coin}, shows that the extra dimension can solely alleviate the coincidence problem, it is obvious that the type of the DE component is very important as well. Clearly, the combination of DGP cosmology with ADE has much more influence in ameliorating the coincidence problem than the $\Lambda$DGP model. We should note here that with attention to Eq.(\ref{eos}), in the limit $n\rightarrow\infty$, our model approaches a $\Lambda$DGP model and if, besides that, $\Omega_{r_c}\rightarrow0$ and therefore $r_c\rightarrow\infty$, we then reach to the standard $\Lambda$CDM model.

\section{Conclusion}\label{s:4}

In this article we investigated the stability analysis of the normal branch of the DGP brane-world model with an ADE component to produce late time acceleration. After introducing a set of new variables we transformed our cosmological equations to an autonomous dynamical system and found the critical points related to the matter dominated and DE dominated epochs. We also found that gravitational screening could be a solution but only at infinity. To show that, we compacted our dynamical system via defining a new dynamical variable $w$ instead of $z$. Moreover, we found that extra dimensional effects can convert a future attractor in 4D cosmology into a saddle point, like the case in \cite{Salcedo}.

Furthermore, a quantitative analysis of statefinder parameters showed that the Universe left an unstable state in the past and  approaches a saddle state in the future. Moreover, it indicated that our model is in good agreement with the $\Lambda$CDM model. Finally, we investigated one of the outstanding cosmological problems, i.e., the coincidence problem and found that the normal DGP with ADE could significantly alleviate this problem.

\acknowledgments
This work has been supported by the grant (SC96PH7686) of Vali-e-Asr University of Rafsanjan, Iran.

\end{document}